\documentclass[journal]{IEEEtran}

\ifCLASSINFOpdf
\else
\fi

\usepackage[numbers, sort&compress]{natbib}
\usepackage{graphicx}
\usepackage{array}
\usepackage{makecell}
\usepackage{comment}
\usepackage{color}

\begin{document}

\title{Semantics-Division Duplexing: A Novel Full-Duplex Paradigm}

\author{Kai~Niu,~\IEEEmembership{Member,~IEEE}, Zijian~Liang,~\IEEEmembership{Graduate Student Member,~IEEE}, Chao~Dong,~\IEEEmembership{Member,~IEEE}, Jincheng~Dai,~\IEEEmembership{Member,~IEEE}, Zhongwei~Si,~\IEEEmembership{Member,~IEEE}, and Ping~Zhang,~\IEEEmembership{Fellow,~IEEE}

\thanks{Corresponding Author: \emph{Kai Niu}. E-mail: niukai@bupt.edu.cn. }
\thanks{The authors are with Beijing University of Posts and Telecommunications. }
}

\maketitle

\begin{abstract}
In-band full-duplex (IBFD) is a theoretically effective solution to increase the overall throughput for the future wireless communications system by enabling transmission and reception over the same time-frequency resources. However, reliable source reconstruction remains a great challenge in the practical IBFD systems due to the non-ideal elimination of the self-interference and the inherent limitations of the separate source and channel coding methods.
On the other hand, artificial intelligence-enabled semantic communication can provide a viable direction for the optimization of the IBFD system.
This article introduces a novel IBFD paradigm with the guidance of semantic communication called semantics-division duplexing (SDD). It utilizes semantic domain processing to further suppress self-interference, distinguish the expected semantic information, and recover the desired sources. Further integration of the digital and semantic domain processing can be implemented so as to achieve intelligent and concise communications. We present the advantages of the SDD paradigm with theoretical explanations and provide some visualized results to verify its effectiveness.

\end{abstract}

\IEEEpeerreviewmaketitle

\section{Introduction}\label{section1}

Duplex technology plays a vital role in supporting bidirectional communications in wireless communication systems. In most of the existing commercial wireless communication systems, including the fifth-generation new radio (5G NR), the frequency division duplex (FDD) or the time division duplex (TDD) technology is relied on to support bidirectional communications \cite{sabharwal2014band, kolodziej2019band}. By dividing transmission resources for uplink and downlink in the frequency or time domain, and incorporating guard bands or periods, FDD or TDD can achieve non-interfering two-way information transmission on wireless resources.

Although widely adopted, these half-duplex (or called out-of-band full-duplex, i.e., OBFD) schemes, have inherent rate limitations in their duplex systems.  This stems from FDD or TDD schemes dividing available time-frequency resources for uplink and downlink transmission without allowing any overlap. Theoretically, ideal in-band full-duplex (IBFD) systems can address these rate limitations. It enables simultaneous uplink and downlink transmissions over the same wireless time-frequency resources, which doubles the spectral efficiency of the full-duplex wireless communication systems \cite{shannon1961two, cover1999elements}.

However, the practical IBFD systems confront considerable challenges in ensuring reliable source reconstruction. From the viewpoint of full-duplex system design, these challenges are attributed to the non-ideal elimination of self-interference. In the practical IBFD systems, the formidable self-interference signals are introduced by the leaked and reflected interference from the transceiver's transmission link; meanwhile, they are also affected by the nonlinear and memory properties due to the employment of non-ideal hardware units \cite{sabharwal2014band, kolodziej2019band}. To achieve effective operation, self-interference cancellation (SIC) methods must be employed to eliminate the self-interference as much as possible from the propagation domain, the frequency domain, and the digital domain \cite{kolodziej2019band}. Unfortunately, these cancellations cannot be ideal, meaning that residual self-interference always exists. Therefore, the follow-up decoding operations are unavoidably impacted by the residual self-interference.

From the viewpoint of coding design, the commonly-used source and channel coding method in these IBFD systems cannot provide a reliable and robust source reconstruction with interference-existing scenarios, resulting in intolerable error reconstructions or even no effective message obtaining. It is due to its inherent limitation, i.e., the decoders are all designed based on the source-channel separation theorem, which is theoretically suboptimal in interference-existing scenarios like the IBFD system due to the absence of an ideal point-to-point condition \cite{cover1999elements, vembu1995source}. This limitation has been extensively discussed and highlighted in classical information theory. Consequently, this inherent limitation indicates that reliable source reconstruction can only be achieved when the received signal power is sufficiently high or when the cell area is sufficiently small, thereby imposing the non-ideal application of the existing IBFD system.

\begin{figure*}[!ht]
\centering
\includegraphics[width=0.95\textwidth]{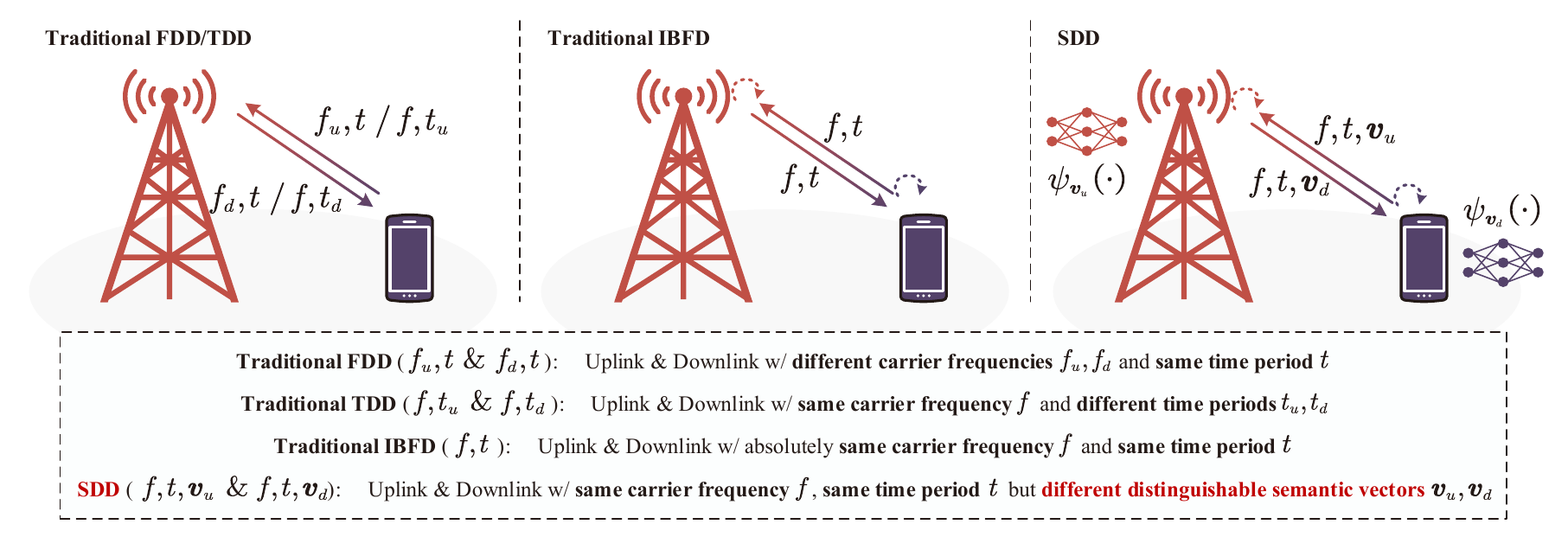}
\caption{The evolution of the full-duplex systems, from the traditional FDD/TDD paradigm to the future-oriented SDD paradigm.}
\label{fig1}
\end{figure*}

This article is devoted to solving the reliable source reconstruction problem of the IBFD systems with the development trend of future wireless communications. Predictably, intelligence and conciseness will be indispensable factors in future wireless communication systems, promoting optimization and transformative changes \cite{zhang2022toward}. Semantic communication, as an artificial intelligence (AI)-enabled novel communication paradigm, offers a promising direction \cite{niu2022paradigm, dai2022communication} mainly based on the joint source-channel coding (JSCC) framework with theoretically optimal performance. Recent works \cite{farsad2018deep, bourtsoulatze2019deep, dai2022nonlinear} mainly focus on point-to-point semantic communications across diverse source modalities, such as text, speech, images, videos, 3D point clouds, and so on. On another line, the exploration of semantic communication in interference-existing scenarios remains a captivating subject of ongoing research, with notable outcomes such as model division multiple-access (MDMA) \cite{zhang2023model}. Both research directions have demonstrated that semantic communication schemes can outperform traditional coding methods.

In this article, we introduce a novel IBFD paradigm called \emph{semantics-division duplexing} (SDD) for future wireless communication systems, which distinguishes the uplink and downlink semantic vectors in high-dimensional semantic spaces. This paradigm enhances traditional IBFD with semantic domain processing and overcomes the reliable source reconstruction problem with the guidance of semantic communications. Inspired by MDMA \cite{zhang2023model}, the processing in the semantic domain can further suppress self-interference, distinguish expected semantic information, and recover desired sources. Integrating digital and semantic domain processing with neural networks, the implementation of SDD can be more intelligent and concise. Under this paradigm, the in-band full-duplex system can achieve bidirectional reliable source reconstruction, improve system service capability, and bolster potential utilization for future wireless communication systems.

Next, we present an overview of the SDD paradigm by introducing the fundamental principles of its realization and outlining its design architecture. We also give the advantages of the SDD paradigm with brief theoretical explanations and provide visualized results to confirm its effectiveness.

\section{Fundamental Principles}\label{section2}

As a novel IBFD paradigm for future wireless communication networks, the design of SDD is expected to aim to achieve the capability of an ideal IBFD system with the help of AI-enabled semantic communications. That is, to provide bidirectional reliable communications with optimal use of available wireless resources. Undoubtedly, the evolution of existing full-duplex paradigms can be the guidance for the design of SDD. Figure \ref{fig1} shows the application scenarios for the traditional FDD/TDD, traditional IBFD, and the future-oriented SDD, illustrating the evolution of the full-duplex paradigms.

The traditional FDD/TDD paradigm allocates available wireless communication resources from the frequency or time dimension for bidirectional transmissions. In other words, the uplink and downlink transmission links can be divided by carrier frequencies or time periods. With no overlapping between uplink and downlink transmission resources, overall transmission rates are limited. Nevertheless, the full-duplex system can be degraded into two independent point-to-point transmission links under this condition, thus providing a relatively reliable source reconstruction based on the source and channel separate coding framework.

The traditional IBFD paradigm implements bidirectional transmissions on the same available wireless time-frequency resources, thus allowing overall transmission rates to be designed higher than traditional FDD/TDD solutions through effective SICs. However, the ideal point-to-point communication conditions are disrupted, manifested by the absolute presence of residual self-interference, which results in the failure to guarantee reliable source reconstruction.

The future-oriented SDD paradigm is required to be designed to approximate the transmission effect of the ideal IBFD system, leveraging the technique and performance advantages over the traditional FDD/TDD and the traditional IBFD paradigm. The fundamental principles of the SDD paradigm can be outlined as follows:

\begin{itemize}
    \item \emph{Inheriting the design method of IBFD:} To make optimal use of available wireless resources, bidirectional transmissions still require to rely on the same available wireless time-frequency resources, and effective SICs need to be also employed to mitigate the powerful self-interference as much as possible.

    \item \emph{Utilizing semantic communication models:} To achieve reliable source reconstruction under absolute interference-existing scenarios, JSCC should be considered a fundamental coding framework \cite{cover1999elements, vembu1995source} that can simultaneously fit coding with source features and channel characteristics. With the help of AI, JSCC-based semantic communication models facilitate the division of distinguishable semantic vectors from the received signals with residual self-interference, and further reliably recover the desired sources.
\end{itemize}

\begin{figure*}[!ht]
\centering
\includegraphics[width=0.97\textwidth]{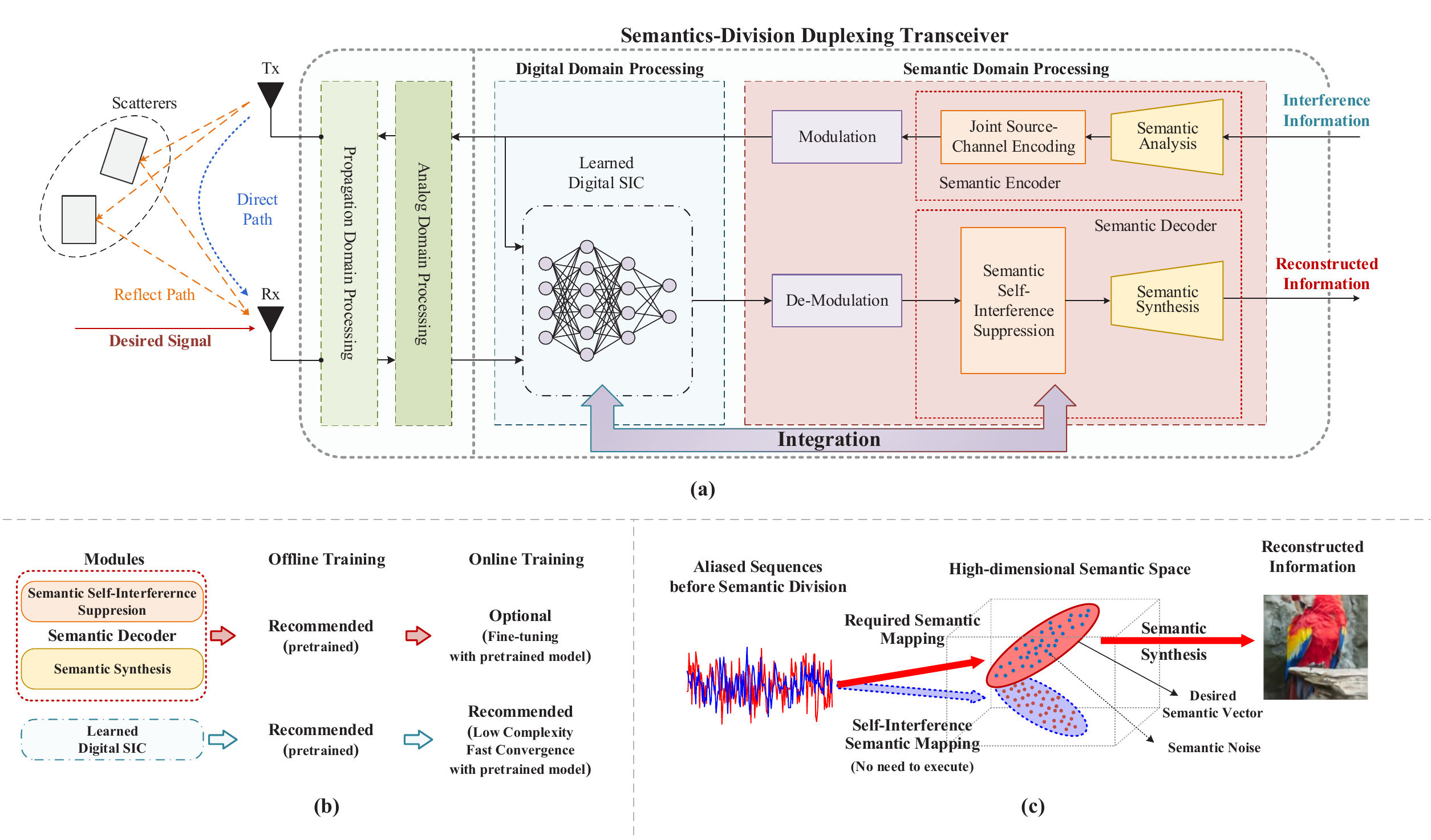}
\caption{The illustration of the SDD transceiver's architecture with its training and semantics-division mechanisms. (a) The architecture of an SDD transceiver. (b) The training mechanisms of the neural network modules inside the SDD transceiver. (c) The semantics-division mechanism used in the Semantic Self-Interference Suppression Block of the SDD Semantic Decoder.}
\label{fig2}
\end{figure*}

\section{Design architecture}\label{section3}

Consider a future scenario where both ends of wireless communication links can support bidirectional SDD transmission. According to the above fundamental principles of the SDD paradigm, in this scenario, both ends use SDD transceivers, enabling transmitted and received signals to share the same wireless time-frequency resources. Therefore, the signal transmitted at one end must serve as the desired received signal at the other end and also function as the self-interference signal at its own end.

To support the SDD bidirectional transmission, the SDD transceiver is required to set a series of processing mechanisms both for the transmitting process and the receiving process:

\begin{enumerate}
    \item In the transmitting process, the sources are first encoded to the semantic signals by the semantic encoder and do the follow-up processing in the digital domain. After power amplification in the analog domain, the semantic signals are sent into the channels by the transmit antennas in the propagation domain.

    \item In the receiving process, a series of SICs are first employed to mitigate the self-interference as much as possible. These cancellations should be processed in the domains same as the existing IBFD paradigm, including the propagation domain, the analog domain, and the digital domain. Then, a semantic decoder is employed to further suppress the self-interference, distinguish the expected semantic information, and recover the desired sources.
\end{enumerate}

Note that semantic information is the high-dimensional features extracted from the sources. It can be processed by the semantic encoder and the decoder based on the high-dimensional semantic spaces instead of the traditional digital signal spaces. In view of this, we name the new domain to process the semantic information as the \emph{semantic domain}, distinguishing it from the digital domain. Therefore, the design mechanism of the SDD transceiver can be described as a four-stage self-interference suppression mechanism, successively mitigating or suppressing self-interference in the propagation domain, the analog domain, the digital domain, and the semantic domain to achieve reliable source reconstruction.

Based on the above-mentioned processing mechanisms, we provide a complete interpretation of the SDD transceiver architecture and its implementation techniques. Figure \ref{fig2} illustrates the SDD transceiver's architecture together with its training and semantics-division mechanisms for reference.

\subsection{The Four-stage Processing Domain}

Figure \ref{fig2}(a) shows an architecture of the designed SDD transceiver. According to the implementation methods in the transceiver, the four-stage processing domains are divided into two categories: the propagation and analog domains are considered the first category, while the digital and semantic domains are regarded as the second category. Herein, we will describe these two categories of processing procedures from the perspective of the receiving process.

The first category includes the processes in the propagation domain and the analog domain, implemented using analog circuits and corresponding settings to resist powerful self-interference:

\begin{itemize}
    \item \emph{For the propagation domain:} Separate-antenna modes can be set to improve the isolation, and both passive and active approaches can be employed to reduce or suppress the leak interference from the direct path \cite{kolodziej2019band}.

    \item \emph{For the analog domain:} Different SICs can be performed in radio-frequency (RF), based-band (BB), and intermediate-frequency (IF) cancellation, as seen in some works, to mitigate analog self-interference \cite{kolodziej2019band}.
\end{itemize}

For the last part of the analog domain processing, the analog signals need to be converted to digital signals with an analog-to-digital converter (ADC), and then provided to the following processing of the second category.

The second category contains the processes in the digital domain and the semantic domain. These two domain processes are separated from the first category because they can both be implemented in processors like digital signal processors (DSP), central processing units (CPU), or graphic processing units (GPU) by specific self-interference algorithms and semantic models:

\begin{itemize}
    \item \emph{For the digital domain:} Linear or nonlinear cancellation algorithms need to be applied to mitigate the self-interference in the digital signals after analog domain processing \cite{sabharwal2014band}. Learned Digital SIC algorithms are more recommended.

    \item \emph{For the semantic domain:} Employing semantic communication models is crucial for further suppressing residual self-interference, distinguishing expected semantic information, and recovering desired sources.
\end{itemize}

Herein, we emphasize that semantic domain processing does not belong to digital domain processing, since the processing mechanisms of these two domains are different:
\begin{itemize}
    \item Digital domain processing conducts \emph{local self-interference cancellations} in traditional signal spaces that can be observed in the digital domain. Conversely, it cannot counteract deeper self-interference that cannot be observed in the digital domain.
    \item Semantic domain processing achieves \emph{global self-interference suppressions} in high-dimensional semantic spaces. It plays a crucial role in suppressing deeper self-interference and ensuring end-to-end transmission performance.
\end{itemize}

\subsection{Semantic Communication Models}

As the most critical difference between the future-oriented SDD and traditional IBFD schemes, semantic communication model-based semantic domain processing is employed to further suppress self-interference, distinguish expected semantic information, and reliably recover desired sources. It is suggested to be designed based on the theoretically optimal JSCC framework \cite{cover1999elements, vembu1995source}, which is simultaneously and jointly optimized according to the semantic features of the sources and the statistic features of the IBFD channels. In other words, semantic models need to be end-to-end optimized in IBFD transmission scenarios. As shown in Fig.~\ref{fig2}(b), the semantic models (semantic decoder and its corresponding semantic encoder) are recommended to be trained offline and deployed online. Further fine-tuning is optional cause the JSCC-based model has a certain capability of error resistance.

The box of semantic domain processing in Fig. \ref{fig2} shows the designed architecture of the semantic communication models employed in the SDD paradigm. For the transmitting process both in the remote sending link and the local self-interference link, a semantic encoder together with a modulation module is placed in the semantic domain processing. Correspondingly, for the receiving process in the local receiving link, a de-modulation module cascaded by a semantic decoder is employed in the semantic domain processing.

For the sending link, the semantic encoder can be decomposed into a concatenation of a semantic analysis block and a joint source-channel encoding (JSCE) block. Specifically, the semantic analysis block extracts the semantic features from the source information, and the JSCE block encodes the semantic features with joint source-channel coding models (e.g., the deep JSCC encoding networks). Certainly, the semantic encoder can be designed and optimized with a single block that simultaneously analyzes and encodes the semantic features. Afterward, the modulation module is utilized to convert the encoded semantic features to the transmitted digital signals, which can be implemented using complex-valued modulation, for example.

Correspondingly, for the receiving link, a de-modulation module is first utilized to perform the reverse operation of the modulation module. Afterward, the semantic decoder processes are undertaken, which can be decomposed into a cascade of a semantic self-interference suppression block and a semantic synthesis block. Specifically, the semantic self-interference suppression block is employed to further suppress residual self-interference in the semantic domain and reconstruct the semantic features by distinguishing the expected semantic information, implemented by a joint source-channel decoding (JSCD) model. Afterward, the semantic synthesis block is employed to recover the desired sources from the reconstructed semantic features. Similarly, the semantic decoder can be designed and optimized within a single block that simultaneously suppresses residual self-interference and recovers the desired sources.

\subsection{Distinguishable Semantic Vectors}

As mentioned above, self-interference cannot be completely eliminated by the series of SIC processes before the semantic domain. Consequently, the received signals of SDD contain bidirectional semantic information including the desired semantic information and a portion of self-interference semantic information. It is different from the situation of the received signals of the traditional IBFD systems, as its coding methods only consider the statistical probabilities of the transmitted symbol sequences instead of the semantic features of the sources.

In order to achieve reliable source reconstruction, a mechanism is required to effectively distinguish the transmitted semantic information from the self-interference semantic information in the semantic self-interference suppression block. As the JSCC-based semantic communication models process information in high-dimensional semantic spaces, it is evident that the semantic vectors carrying different semantic information are distinguishable in some specific high-dimensional spaces that can be processed based on semantic decoders. Therefore, the distinguishable semantic vectors have the potential to help the SDD systems resolve the bidirectional transmitted signals.

Clearly, to recognize different semantic vectors from time-frequency aliased signals, it is essential to utilize different semantic mappings with the semantic decoder for these aliased signals in different directions. As shown in Fig.~\ref{fig2}(c), the aliased received signals in SDD can be mapped to different semantic vectors using diverse semantic mappings, allowing the recognition of the desired received signal with its corresponding semantic mappings.

However, the effect of this semantic division may not be ideal: the desired received signal may not be entirely distinguished by its semantic mappings with the residual self-interference. Since the input signal of the semantic mappings is already affected by aliased self-interference, the residual self-interference will be suppressed into semantic noise on the desired semantic vectors, shown as the dots on desired semantic features in Fig.~\ref{fig2}(c). By optimizing the semantic mappings corresponding to the desired received signals, the influence of the semantic noise will be effectively weakened, allowing for high-accuracy recognition of the required semantic information in the processing of the semantic domain.

In view of this, the semantic self-interference suppression block of the semantic decoder shown in Fig. \ref{fig2} is expected to distinguish the corresponding desired semantic vectors. This JSCD-based model is required to be trained to achieve diverse semantic mappings and recognize their receiving semantic information under various IBFD scenarios like two-way communications. To achieve this goal, a simple approach is to train different JSCD-based models for targeted distinguishing bidirectional semantic vectors. However, it will result in increased storage challenges for the models as the IBFD scenario becomes more complex. As a workaround, a single JSCD-based model can also be trained to recognize different semantic information by providing different conditions, which are additionally transmitted by the semantic encoders as the side information. This approach can effectively distinguish different semantic vectors and address the storage problem for the semantic models while incurring only a minimal increase in bandwidth overhead.

\subsection{Integration for Intelligence and Conciseness}

As mentioned before, the implementation of the propagation and the analog domain relies on the implementation of analog circuits, yet the processing of the digital and the semantic domains can be achieved with algorithms and models on the same computing processors. In order to achieve an intelligent and concise design, these algorithms and models need to be designed, implemented, and optimized based on integrated deep neural networks. This can be achieved by incorporating the digital SIC processing employed in a nonlinear manner \cite{kolodziej2019band}, i.e., Learned Digital SIC.

As stated in Fig.~\ref{fig2}(b), the Learned Digital SIC is recommended to be first trained offline, using various types of sources coded sequences, and then to be online fine-tuned after deployment. Due to the strong data fitting ability of the nonlinear structure of the neural network, the learned digital SIC method is expected to converge faster, have lower processing complexity, and consume less energy.

The architecture in Fig. \ref{fig2}(a) shows an integration of the digital domain processing and the semantic domain processing. In this architecture, learned digital SIC based on neural networks are integrated with the semantic decoder, incorporating the processing and optimization of the digital domain and the semantic domain process for intelligence and conciseness.

Predictably, the development trend of future wireless communication systems toward intelligence and conciseness will extend to the intelligent combination of hardware and software. When the time comes, SDD will have the ability to optimize the analog circuit structure of the propagation domain and the analog domain using AI-enabled deep neural networks and optimization mechanisms, achieving complete integration with all processing domains.

\begin{figure*}[!ht]
\centering
\includegraphics[width=1.0\textwidth]{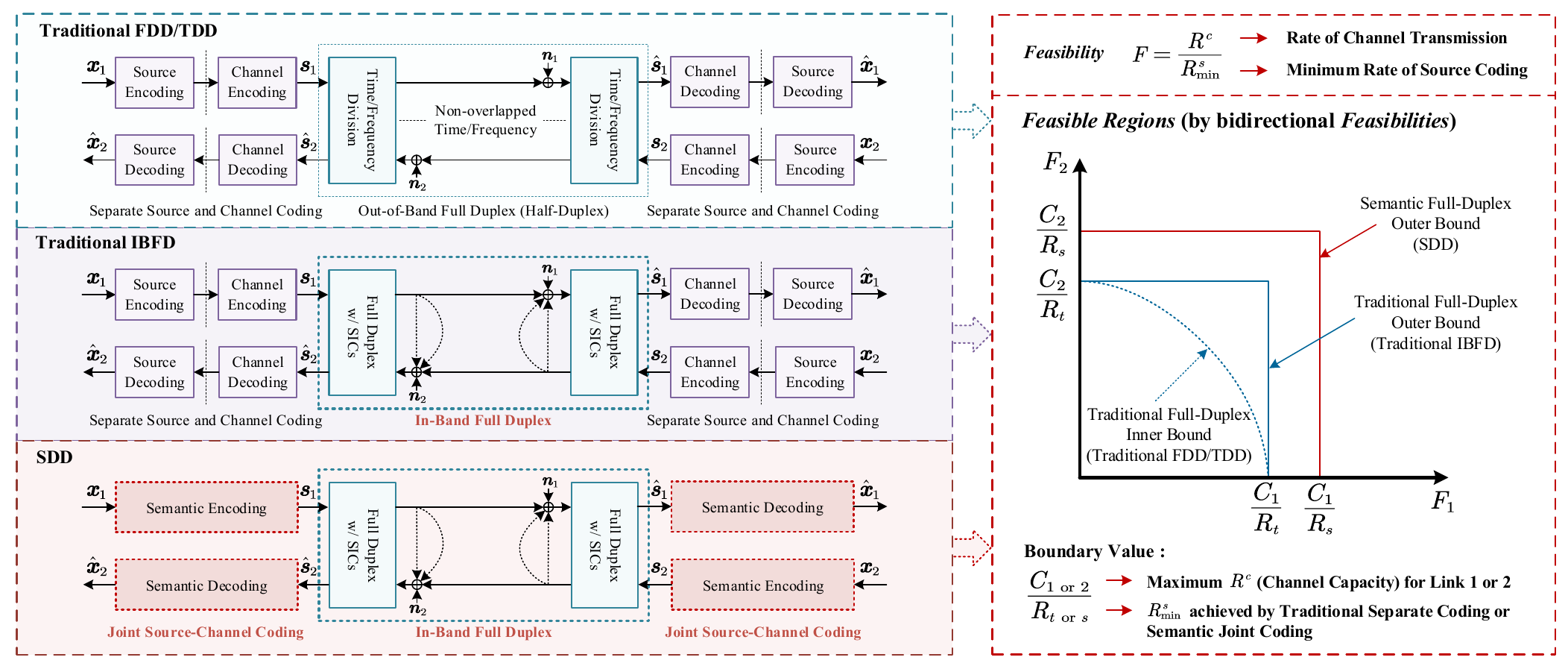}
\caption{The duplex and coding methods of different full-duplex paradigms with their theoretical \emph{Feasible Regions}.}
\label{fig4}
\end{figure*}

\section{Theoretical Explanations}\label{section4}

From the perspective of information theory, the optimal performance of full-duplex systems can be evaluated by the rate regions of the two-way communication. Based on Shannon's research on the two-way transmission model \cite{shannon1961two}, an inner bound and an outer bound can be derived to describe the optimal rate regions for different full-duplex paradigms. The inner bound represents the optimal rate regions when the self-interference is completely independent of the received signals, which corresponds to the traditional FDD/TDD paradigm. The outer bound stands for the optimal rate regions when the self-interference can be entirely eliminated from the received signals, guiding the design of the traditional IBFD paradigm.

However, such rate regions cannot evaluate the differences between the traditional IBFD paradigm and the SDD paradigm. The most obvious evidence is that even with the same channel capacity and the same rate region, SDD still surpasses the traditional IBFD in guaranteeing reliable source reconstruction due to the efficient utilization of the JSCC framework and AI-enabled semantic communication models. This issue arises from the absence of the factors associated with the sources in the capacity-based evaluation method.

Therefore, we introduce a new theoretical metric \emph{Feasibility} and its corresponding region \emph{Feasible Region} in \cite{zhang2023model} to explain the apparently-existing performance differences. The metric \emph{Feasibility} is derived from the source-channel coding theorem and serves to characterize the service capability of a wireless communication system utilizing a specific source and channel coding scheme for a given channel. For a point-to-point wireless transmission system, \emph{Feasibility} can be expressed as the ratio of the channel transmission rates to the source coding rates. The \emph{Feasible Region} is constructed by the bidirectional \emph{Feasibilities}, defined as the ratios of the channel transmission rates in the corresponding direction to the minimum source coding rates between the bidirectional source transmission.

Figure \ref{fig4} shows the optimal \emph{Feasible Regions} of different full-duplex paradigms. From the definition, it is evident that the optimal \emph{Feasible Regions} is related to both the channel capacity and the coding efficiency. Some outcomes can be obtained from this figure:

\begin{enumerate}
    \item The traditional FDD/TDD paradigm achieves the lowest \emph{Feasible Region} inner bound since both the duplexing mechanism and the coding method follow the simplest designing schemes. It performs the out-of-the-band duplex and the separate source and channel coding to support the bidirectional transmission.

    \item The traditional IBFD paradigm exhibits a larger \emph{Feasible Region} outer bound compared to the traditional FDD/TDD paradigm. This outcome is obtained from the improvement of the channel capacity due to the employment of IBFD-based bidirectional channel design, which simultaneously transmits and receives the signals on the same wireless time-frequency resources and employs effective self-interference cancellations.

    \item The SDD paradigm presents another improvement in \emph{Feasible Region} compared to the traditional IBFD paradigm, expanding the outer bound. This advantage is derived from the improvement of the coding efficiency due to the employment of JSCC-based well-designed semantic communication models, which jointly consider the semantic features of the sources and the statistical features of the IBFD channels.
\end{enumerate}

To summarize, the SDD paradigm has the optimal outer bound of \emph{Feasible Regions} due to its IBFD-based channel design and the JSCC-based coding design. This implies that up to now, SDD is the theoretically optimal full-duplex paradigm that can provide the highest-level service capability for the system. Therefore, with proper design and optimization, SDD has the potential to outperform all the existing full-duplex systems in terms of overall performance.

\begin{figure}[!t]
    \centering
    \includegraphics[width=0.45\textwidth]{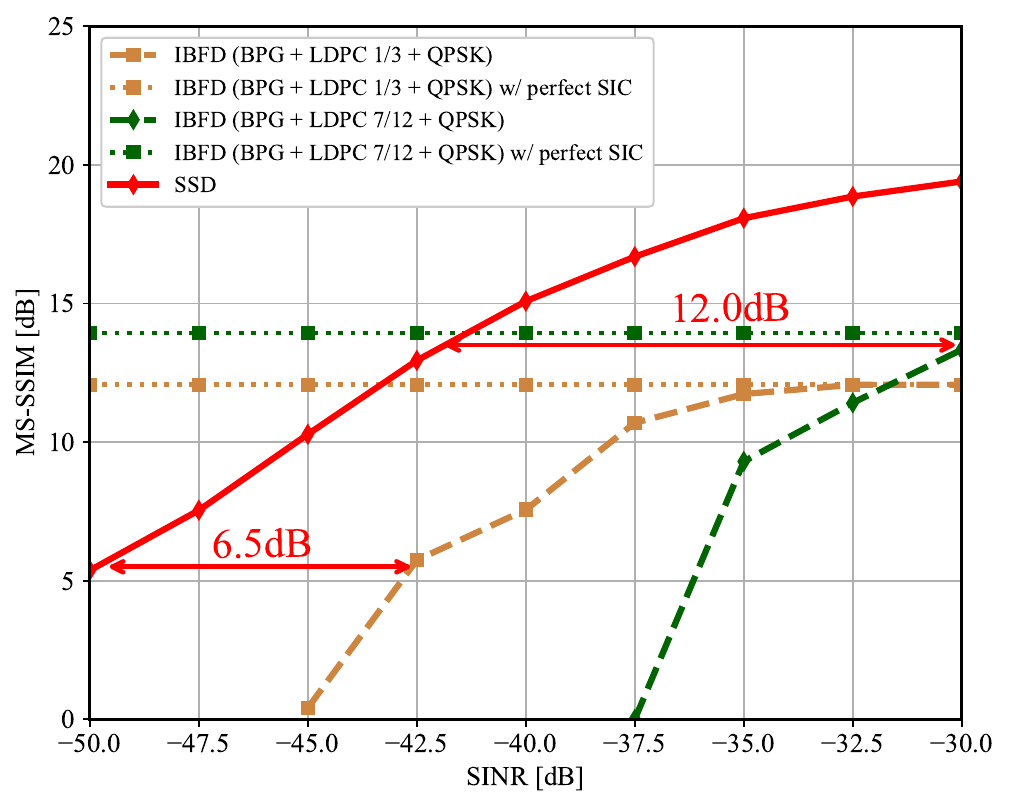}
    \caption{Example of the MS-SSIM performance achieved by different full-duplex demos for the Rician-distributed self-interference scenario.}
    \label{fig5}
\end{figure}

\begin{figure*}[t]
\centering
\includegraphics[width=0.91\textwidth]{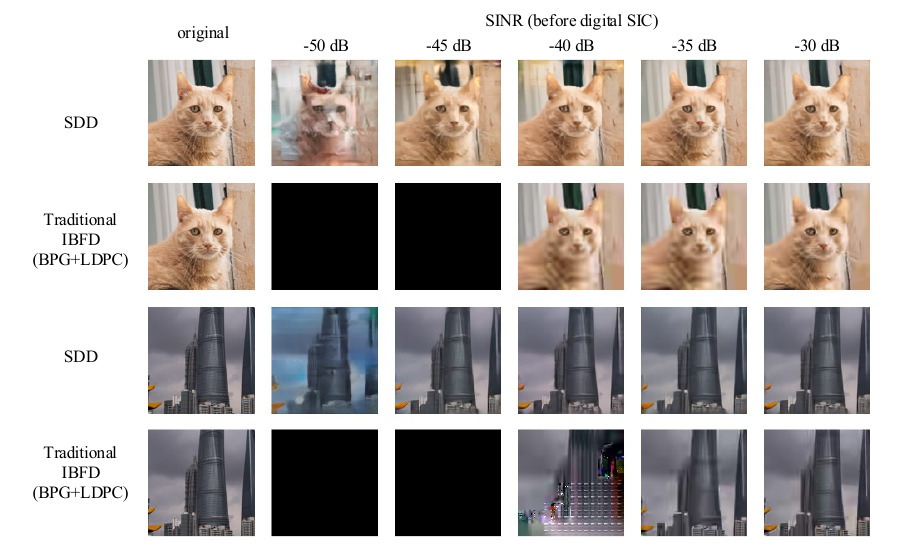}
\caption{Example of visual comparisons within the SINR range of --50dB to --30dB for the Rician-distributed self-interference scenario, in which the black images represent the images that cannot be obtained by the BPG decoder at the corresponding SINR.}
\label{fig6}
\end{figure*}

\section{Implementations Showcase}\label{section5}



As indicated in Fig. \ref{fig2}, implementing the SDD transceiver architecture is a challenging issue. However, a simplified version of SDD can be implemented based on the existing IBFD systems. That is, replace the traditional source and channel coding methods with the anti-interference semantic communication models, and change the digital SIC algorithms with neural network-based nonlinear digital SIC methods within these traditional IBFD systems. This implementation method is compatible with the existing systems, including self-interference cancellation processes, thereby lowering the difficulty of implementation relatively.

We implement a demo program to simulate a portion of the SDD transceiver processes, including digital domain processing and semantic domain processing. The demo is performed under a two-way image transmission scenario, in which the images are resized and cropped to 128 $\times$ 128 patches from the OpenImages datasets \cite{kuznetsova2020open}. A neural network-aided digital SIC method \cite{kong2022neural} is employed in the digital domain, and a non-linear transform source-channel coding model \cite{dai2022nonlinear} already-trained under the interference-existing circumstances is set for the semantic domain both in the sending/self-interference link and the receiving link. The evaluation metric is the multi-scale structural similarity (MS-SSIM) between the reconstructed sources and the sent sources.

For comparison, a traditional IBFD demo is also implemented and simulated. The same digital SIC method with the SDD demo is employed in the digital domain. The image codec \emph{Better Portable Graphics} (BPG), the 5G NR low-density parity check (LDPC) codes with the first base graph (BG1), and QPSK modulation are employed to implement the traditional source and channel coding method for both the sending/self-interference link and the receiving link. With fairness as the guiding principle, both SDD and the comparison scheme set the same bandwidth overhead with their employed coding methods. According to this principle, the quality parameter (QP) of BPG is set to 37 when the coding rates of LDPC are 7/12, and 41 when the coding rates of LDPC are 1/3.

Considering the transmitting mode of the self-interference signals, we model the propagation domain and the analog domain processing as a Rician-distributed multi-path invariant channel system for the testing demos. A 5G NR path loss model is considered for the transmitted signals with 2.9 GHz carrier frequency \cite{sun2016propagation}, and the distance between the transceivers is controlled to 25$\sim$50 meters. The delay and the received power ratio of this multi-path channel refer to the configuration in \cite{kong2022neural}, but the power of the received self-interference signals are set to simulate the signal-to-interference-plus-noise (SINR) before the digital domain processing across the range of --50$\sim$--30 dB.

We plot the simulation results in Fig. \ref{fig5}. It shows that SDD performs a graceful quality degradation as the SINR decreases, demonstrating its reliability. Additionally, compared with traditional IBFD schemes, a 6.5$\sim$12.0 dB performance gain can be observed from the metric of MS-SSIM across the range of 5.5$\sim$13.5 dB. Compared with traditional IBFD schemes with perfect SIC, SDD still maintains advantages in the higher 12.0 dB range to achieve better reconstruction quality. This result indicates the effectiveness of the SDD paradigm. From the perspective of implementation effects, these advantages reveal an outcome that an SDD system can be employed in cells with greater areas than a traditional IBFD system, thereby effectively reducing the difficulties of IBFD systems in practical implementations.

Figure \ref{fig6} shows an example of visual comparisons derived from the simulations of the above demos. In these results, the traditional IBFD exhibits significant instability. Within the lower range of SINRs, it introduces significant errors in the reconstructed images or even fails to reconstruct the desired source images. On the contrary, the reconstructed images produced by SDD maintain the reliability of important semantic features even in the presence of certain errors at lower SINR ranges, further demonstrating the robustness of the SDD paradigm.

To summarize, the implementation of the employed SDD demo can enhance the reliability, effectiveness, and robustness of the IBFD systems. With effective implementation, SDD-based full-duplex equipment has the potential to greatly improve the overall performance of future wireless communication systems.

The implementation of the future SDD architecture requires further optimization to meet the intelligent and concise requirements of future wireless communication networks. Not only the integration of digital and semantic domain processing but the designs of analog circuits in the propagation domain and the analog domain may also be optimized with AI, thereby achieving a more comprehensive integration.

\section{Conclusions and Future Directions}\label{section6}

In this article, we introduced a novel full-duplex paradigm called Semantics-Division Duplexing. The fundamental principles for the SDD paradigm design are first presented, and then a detailed architecture of an SDD transceiver is designed based on these fundamental principles and toward the intelligent and concise requirements of future wireless communication networks. From the theoretical explanation and the implementation showcase, the SDD paradigm can be more reliable, more effective, and more robust than the traditional IBFD paradigm, and has the potential to greatly improve the overall performance of future wireless communication systems. For future works, it is a confident direction to study the optimization theory and design frameworks oriented towards maximizing bidirectional Feasibility metrics to attain the outer bound of the SDD's \emph{Feasible Region}. Furthermore, the study of future SDD systems oriented to large-scale ubiquitous connection scenarios like smart cities and more severe communication scenarios like underwater acoustic communications may be potential and vital directions.

\section*{Acknowledgements}\label{Acknowledgements}

This work was supported by the National Natural Science Foundation of China (No. 62293481, No. 62071058).

\section*{Biographies}\label{Biographies}

\begin{IEEEbiography}[{\includegraphics[width=1in,height=1.25in,clip,keepaspectratio]{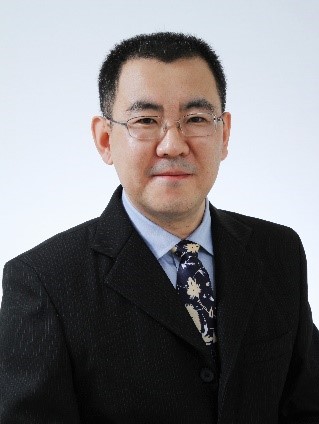}}]{Kai Niu}
    [M'12] (niukai@bupt.edu.cn) received the B.S. degree in information engineering and the Ph.D. degree in signal and information processing from the Beijing University of Posts and Telecommunications (BUPT), Beijing, China, in 1998 and 2003, respectively. He is currently a Professor with the School of Artificial Intelligence, BUPT. His research interests include channel coding theory and applications, semantic communication, and broadband wireless communication.
\end{IEEEbiography}

\begin{IEEEbiography}[{\includegraphics[width=1in,height=1.25in,clip,keepaspectratio]{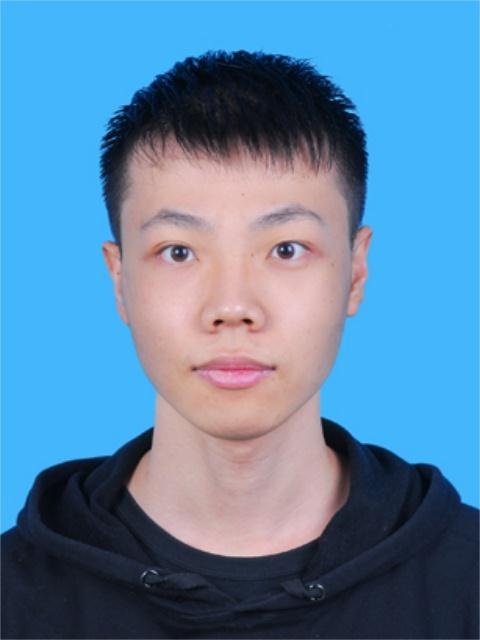}}]{Zijian Liang}
    [S'23] (liang1060279345@bupt.edu.cn) received the B.S. degree in information engineering and the M.Eng. degree in electronics and communication engineering from Beijing University of Posts and Telecommunications (BUPT), Beijing, China, in 2019 and 2022, respectively. He is currently pursuing the Ph.D. degree with the School of Artificial Intelligence, BUPT. His research focuses on wireless communications, source and channel coding, and semantic communication.
\end{IEEEbiography}

\begin{IEEEbiography}[{\includegraphics[width=1in,height=1.25in,clip,keepaspectratio]{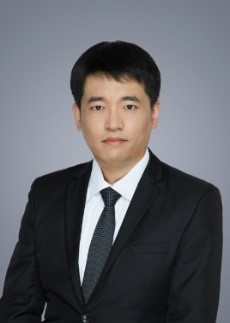}}]{Chao Dong}
    [M'17] (dongchao@bupt.edu.cn) received his B.S. and Ph.D. degrees in signal and information processing from the Beijing University of Posts and Telecommunications (BUPT), Beijing, China, in 2007 and 2012, respectively. He is currently an Associate Professor with the School of Artificial Intelligence, BUPT. His research interests include MIMO signal processing, multiuser precoding, decision feedback equalizer, and relay signal processing.
\end{IEEEbiography}

\begin{IEEEbiography}[{\includegraphics[width=1in,height=1.25in,clip,keepaspectratio]{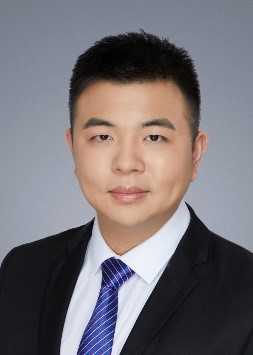}}]{Jincheng Dai}
     [S'17, M'21] (daijincheng@bupt.edu.cn) received the B.S. and Ph.D. degree from the Beijing University of Posts and Telecommunications (BUPT), Beijing, China, in 2014 and 2019. He is currently an Associate Professor with the School of Artificial Intelligence, BUPT. His research focuses on semantic communications, source and channel coding, and machine learning for communications.
\end{IEEEbiography}

\begin{IEEEbiography}[{\includegraphics[width=1in,height=1.25in,clip,keepaspectratio]{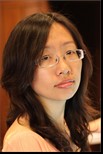}}]{Zhongwei Si}
     [M'16] (sizhongwei@bupt.edu.cn) received the Ph.D. degree from the KTH Royal Institute of Technology, Sweden, in 2013. In 2013, she joined the Beijing University of Posts and Telecommunications, where she is currently an Associate Professor. Her research interests include wireless communication, information theory, and data mining.
\end{IEEEbiography}

\begin{IEEEbiography}[{\includegraphics[width=1in,height=1.25in,clip,keepaspectratio]{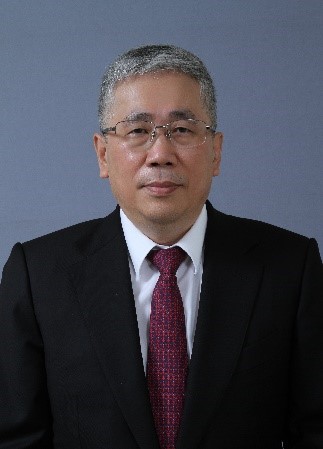}}]{Ping Zhang}
     [M'07, SM'15, F'18] (pzhang@bupt.edu.cn) is currently a Professor of School of Information and Communication Engineering at Beijing University of Posts and Telecommunications, the director of State Key Laboratory of Networking and Switching Technology, a member of IMT-2020 (5G) Experts Panel, a member of Experts Panel for China’s 6G development. He served as Chief Scientist of National Basic Research Program (973 Program), an expert in Information Technology Division of National High-tech R\&D program (863 Program), and a member of Consultant Committee on International Cooperation of National Natural Science Foundation of China. His research interests mainly focus on wireless communication. He is an Academician of the Chinese Academy of Engineering (CAE).
\end{IEEEbiography}

\end{document}